\documentclass[11pt]{article}  
\usepackage{menuproc}
\usepackage{cite}
\usepackage{epsfig}
\usepackage{amsmath,amssymb}
\newcommand {\eq}{$$}
\newcommand {\qe}{$$}
\newcommand {\cen}[1]{\begin{center} #1 \end{center}}
\newcommand {\bfk} {{\bf k}}
\newcommand {\bfq} {{\bf q}}
\newcommand {\bfp} {{\bf p}}
\newcommand {\bfa} {{\bf a}}
\newcommand {\bfb} {{\bf b}}
\newcommand {\h}{\frac{1}{2}}
\newcommand {\slq}{\slash{\!\!\!q}}
\newcommand {\slp}{\slash{\!\!\!p}}

\newcommand {\bfm} {{\bf m}}
\newcommand {\bfn} {{\bf n}}
\newcommand {\bfl} {{\bf l}}

\begin{document}
%
%
%
\titlematter{Two-Pion Exchange in proton-proton Scattering}%
{ W. R. Gibbs$^a$ and Beno\^\i t Loiseau$^b$}%
{$^a$ New Mexico State University, Department of Physics,\\
 $^b$ LPNHE Universit\'e P. \& M. Curie}%

{We present calculations of the two-pion-exchange contribution to
proton-proton scattering at 90$^{\circ}$ using form factors appropriate
for representing the distribution of the constituent partons of the
nucleon.
{\bf Talk given at MENU2001, George Washington University July 26-31,
2001}
}

\section{Introduction}

The cross section and the spin-correlation observable, C$_{\rm NN}$,
measured in the region from 0 to 12 GeV/c P$_{lab}$ at 90$^{o}$ provide an
excellent testing ground for our theories of hadronic interactions. These
data have been the focus of a number of experimental\cite{exps} and
theoretical studies\cite{theo}. There is an apparent simplification of the
amplitude in the region from 4 GeV/c to 8 GeV/c.  At 90 $^{\circ}$ the
spin-correlation observable $ C_{NN}$ has a constant value of about 0.07
which might indicate the dominance of a single mechanism. The simple quark
exchange mechanism gives 1/3 for $C_{NN}$.

One pion exchange provides an important contribution below about 2 GeV/c. 
The exchange of two pions has been used as a basis for the NN interaction
at low energies and it is reasonable to assume that it remains important
at higher energies. We have calculated the box and crossed two-pion
exchange Feynman graphs with nucleons as intermediate states. Even if this
mechanism dominates one can not expect good quantitative agreement with
the data due to the strong inelasticity of the NN interaction at these
energies since distortion effects need to be considered.  None the
less, several questions can be raised and answered. 

\section{Two-pion Exchange Calculation}

The full cross section is given by
$$
\frac{d\sigma}{d\Omega}=\left(\frac{m^2}{4\pi E}\right)^2|M|^2.
$$
where, for pseudo-scalar coupling and box kinematics M is 

\eq M_{PS}=g^4
\frac{1}{(2\pi)^4}\int \frac{dq  
\left[\bar{u}(k')\gamma_5(\slash{\!\!\!p}+m)\gamma_5u(k)\right]_1
\left[\bar{u}(-k')\gamma_5(\slash{\!\!\!p'}+m)\gamma_5u(-k)\right]_2}
{(p^2-m^2+i\epsilon)({p'}^2-m^2+i\epsilon)(q^2-{\mu}^2+i\epsilon)
({q'}^2-{\mu}^2+i\epsilon)}.
\qe
In this expression the minus sign on $k$ and $k'$ applies only to
spatial components.
Corresponding to the propagator for proton 1, we have

$$ 
\bar{u}(k')\gamma_5(\slash{\!\!\!p}+m)\gamma_5u(k)=
\bar{u}(k')\gamma_5[(p_0-E_p)\gamma_0+E_p\gamma_0
-\mbox {\boldmath $\gamma$}\cdot\bfp+m]\gamma_5u(k)$$ 

$$ =-\bar{u}(k')(p_0-E_p)\gamma_0u(k)
+2m\bar{u}(k')\gamma_5\sum_r u_r(p)\bar{u}_r(p)\gamma_5u(k)
$$
where the spinors $u_r(p)$ are on shell with energy
$ E_p\equiv \sqrt{\bfp^2+m^2}. $

The integral can be written as an operator in spin space in the form
\eq M_{PS}=g^4
\frac{1}{(2\pi)^4}\int \frac{dq  \Theta_1(p)\Theta_2(p')}
{(p^2-m^2)({p'}^2-m^2)(q^2-\mu^2)({q'}^2-\mu^2)},
\label{gen}
\qe

$$ {\rm where}\ \ \Theta_1(p) 
=\frac{1}{2m}\left\{\left(E_p-p_0\right)\left(
E+m+\frac{\mbox{\boldmath $\sigma_1$}\cdot\bfk '\mbox{\boldmath $\sigma_1$}\cdot\bfk}{E+m}\right)+
\right. 
$$
\eq \left. \frac{
\left[(E+m)\mbox{\boldmath $\sigma_1$}\cdot\bfp-(E_p+m)\mbox{\boldmath $\sigma_1$}\cdot\bfk '\right]
\left[(E_p+m)\mbox{\boldmath $\sigma_1$}\cdot\bfk-(E+m)\mbox{\boldmath $\sigma_1$}\cdot\bfp\right]}
{(E_p+m)(E+m)} \right\}.
\qe

In the case of pseudo-vector coupling  the interaction is given by
$\frac{f}{\mu}\bar{\psi}\gamma_{\mu}\gamma_5\mbox {\boldmath $\tau$}\cdot
\partial^{\mu}\mbox {\boldmath $\phi$}_{\pi}\psi $ 
and the operator corresponding to the first proton propagator is
\eq
(p^2-m^2)P_{PV}=-\frac{f^2}{\mu^2}\bar{u}(k')
\slq'\gamma_5(\slp+m)\slq\gamma_5u(k)
\qe
\eq
=
\frac{f^2}{\mu^2}\bar{u}(k')\gamma_5(p^2-m^2)(\slp+3m)\gamma_5u(k)
+\frac{f^24m^2}{\mu^2}\bar{u}(k')
\gamma_5(\slp+m)\gamma_5u(k).
\qe

\begin{figure}[t]
\parbox{.4\textwidth}{\epsfig{file=
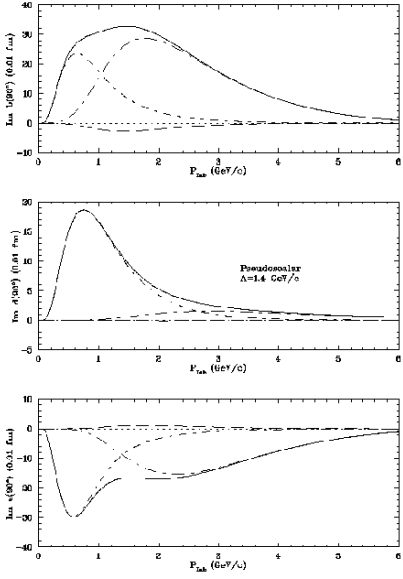,width=.4\textwidth,silent=,clip=}}
\hfill
\parbox{.5\textwidth}{\caption{\label{sidebyside}
Contributions to the imaginary part of the b, d and e pseudo-scalar
amplitudes. The short dash-dot line shows the result of placing the two
nucleons on shell. The dashed and long dash-dot curves show the principal
value parts from the box and cross diagrams respectively. The
two-nucleons-on-shell contribution dominates at low energy but dies out at
high energy where the amplitudes are dominated by the principal-value part
of the crossed diagram.}}
\end{figure}
The operator naturally separates into a term which corresponds to a
contact term and one which is identical to the pseudo-scalar expression
given before

\eq P_{PV}=\frac{f^2}{\mu^2}\bar{u}(k') 
\gamma_5(\slp+3m)\gamma_5u(k) 
+g^2\frac{\bar{u}(k')\gamma_5(\slp+m)\gamma_5u(k)}
{p^2-m^2}=P_C+P_{PS}.
\qe

Defining an operator analogous to $\Theta_1(p)$
for the contact term for the first proton

$$ C_1(p)
=\frac{E+m}{2m}\left[E_p-p_0+2m
+\left(E_p-p_0-2m\right)
\frac{\mbox{\boldmath $\sigma_1$}\cdot\bfk '\mbox{\boldmath $\sigma_1$}\cdot\bfk}{(E+m)^2}\right]+
$$
$$ \frac{
\left[(E+m)\mbox{\boldmath $\sigma_1$}\cdot\bfp-(E_p+m)\mbox{\boldmath $\sigma_1$}\cdot\bfk '\right]
\left[(E_p+m)\mbox{\boldmath $\sigma_1$}\cdot\bfk-(E+m)\mbox{\boldmath $\sigma_1$}\cdot\bfp\right]}
{2m(E_p+m)(E+m)},
\qe
we can write
$$
M_{PV}=
\frac{g^4}{16m^4}\frac{1}{(2\pi)^4}\int dq\frac{C_1(p)
C_2(p')}{(q_0^2-\omega^2)(q_0^2-{\omega'}^2)}
+\frac{g^4}{4m^2}\frac{1}{(2\pi)^4}\int dq\frac{C_1(p)
\Theta_2(p')}{({p'}^2-m^2)(q_0^2-\omega^2)(q_0^2-{\omega'}^2)}
$$
$$
+\frac{g^4}{4m^2}\frac{1}{(2\pi)^4}\int dq\frac{\Theta_1(p)
C_2(p')}{(p^2-m^2)(q_0^2-\omega^2)(q_0^2-{\omega'}^2)}
+M_{PS}.
\qe

\begin{figure}[t]
\parbox{.4\textwidth}{\epsfig{file=
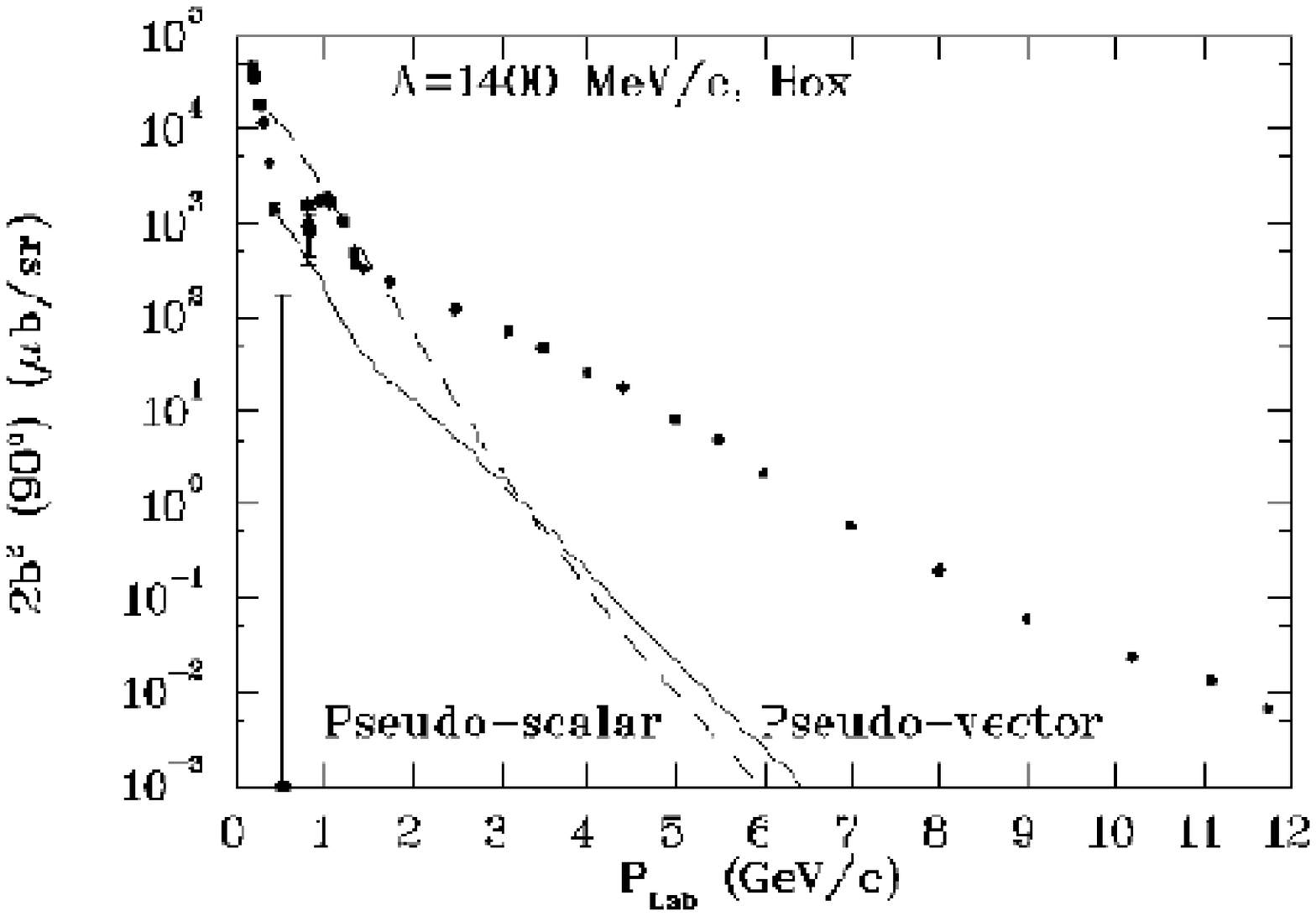,width=.7\textwidth,silent=,clip=}}
\hfill
\parbox{.4\textwidth}{\caption{\label{b2b1400}
The calculation of $2|b|^2$ for the box diagram.
The dashed line shows the result for the pseudo-scalar coupling 
and the solid line that of the pseudo-vector.}}
\end{figure}

\section{Proton Structure}

The expressions in the previous section lack the form factors which are
associated with each pion-nucleon vertex.  We treat the two protons as
particles with intrinsic size related to the distribution of (primarily)
constituent quarks in the nucleon, assuming that the underlying
interaction of the pions is with the partons. The form factor can be
directly obtained as the Fourier transform of the density\cite{nous}. Such
a derivation is inherently non-relativistic since it is not expressed in
terms of Lorentz invariants.

\begin{figure}[t]
\parbox{.4\textwidth}{\epsfig{file=
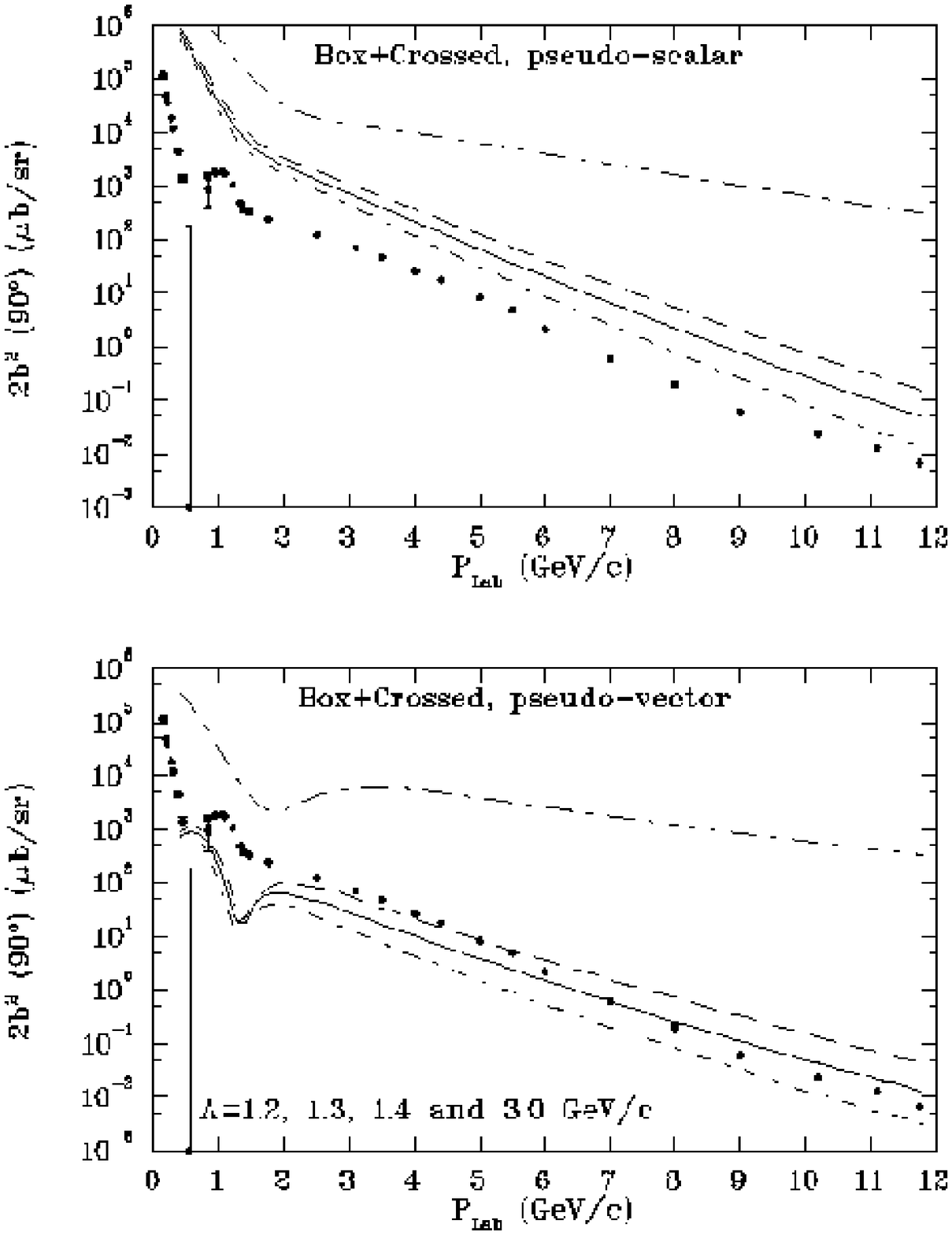,width=.5\textwidth,silent=,clip=}}
\hfill
\parbox{.4\textwidth}{\caption{\label{b2tcmp}
Dependence of $|b|^2$ on the value of $\Lambda$. The smaller values
of the partial cross sections correspond to the smaller values of
$\Lambda$ in order. }}
\end{figure}

An exponential parton density leads to the form,
$(\alpha^2-\mu^2)^2/(\bfq^2+\alpha^2)^2$.
A common relativistic generalization is 
$$
\frac{(\alpha^2-\mu^2)^2}{(q_0^2-\bfq^2-\alpha^2)^2},
$$
\label{mesonform}
but this procedure introduces an additional singularity in $q_0$ on the
real axis.  

\begin{figure}[t]
\parbox{.4\textwidth}{\epsfig{file=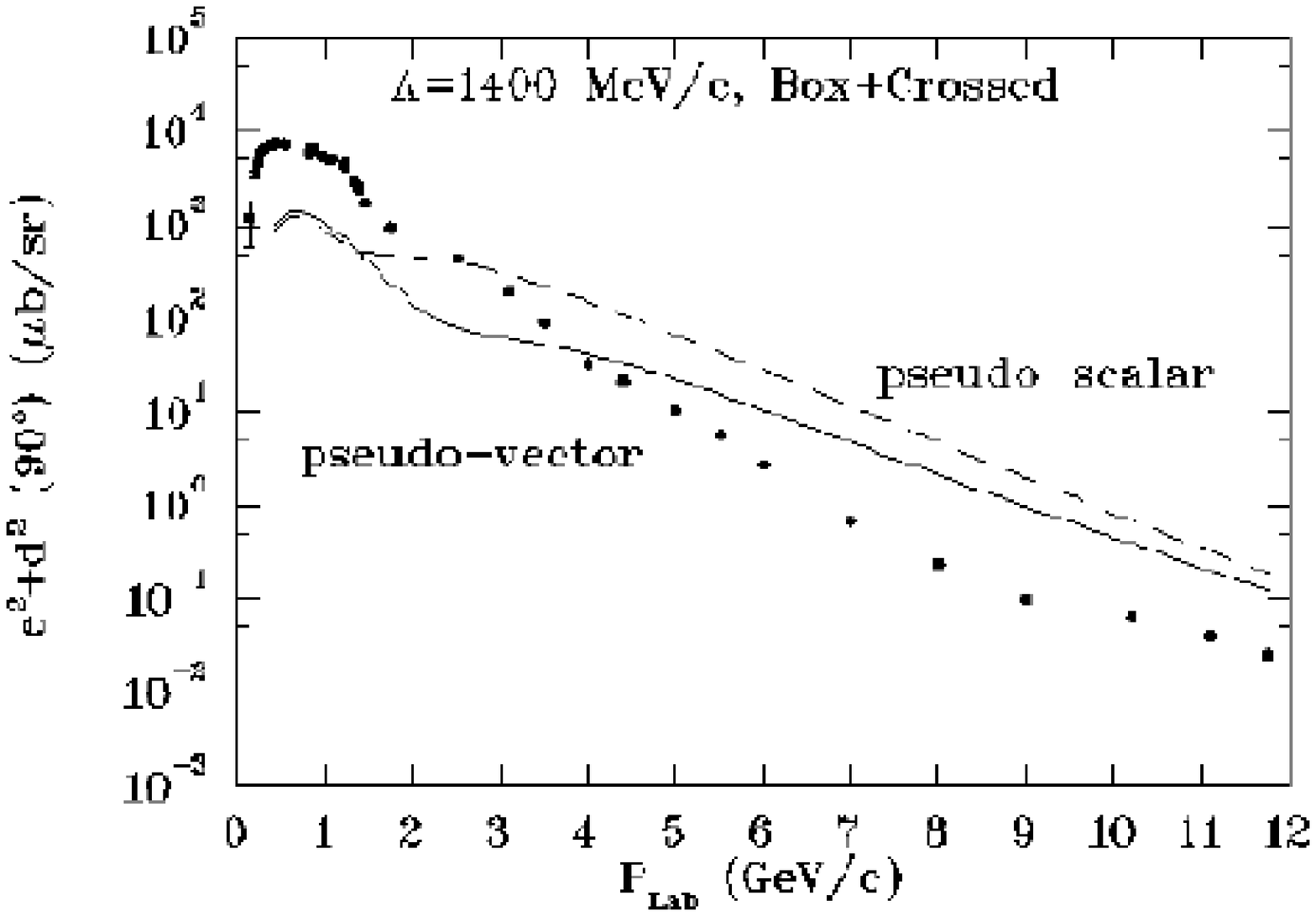,
width=.45\textwidth,silent=,clip=}}
\hspace*{.5in}
\parbox{.4\textwidth}{\epsfig{file=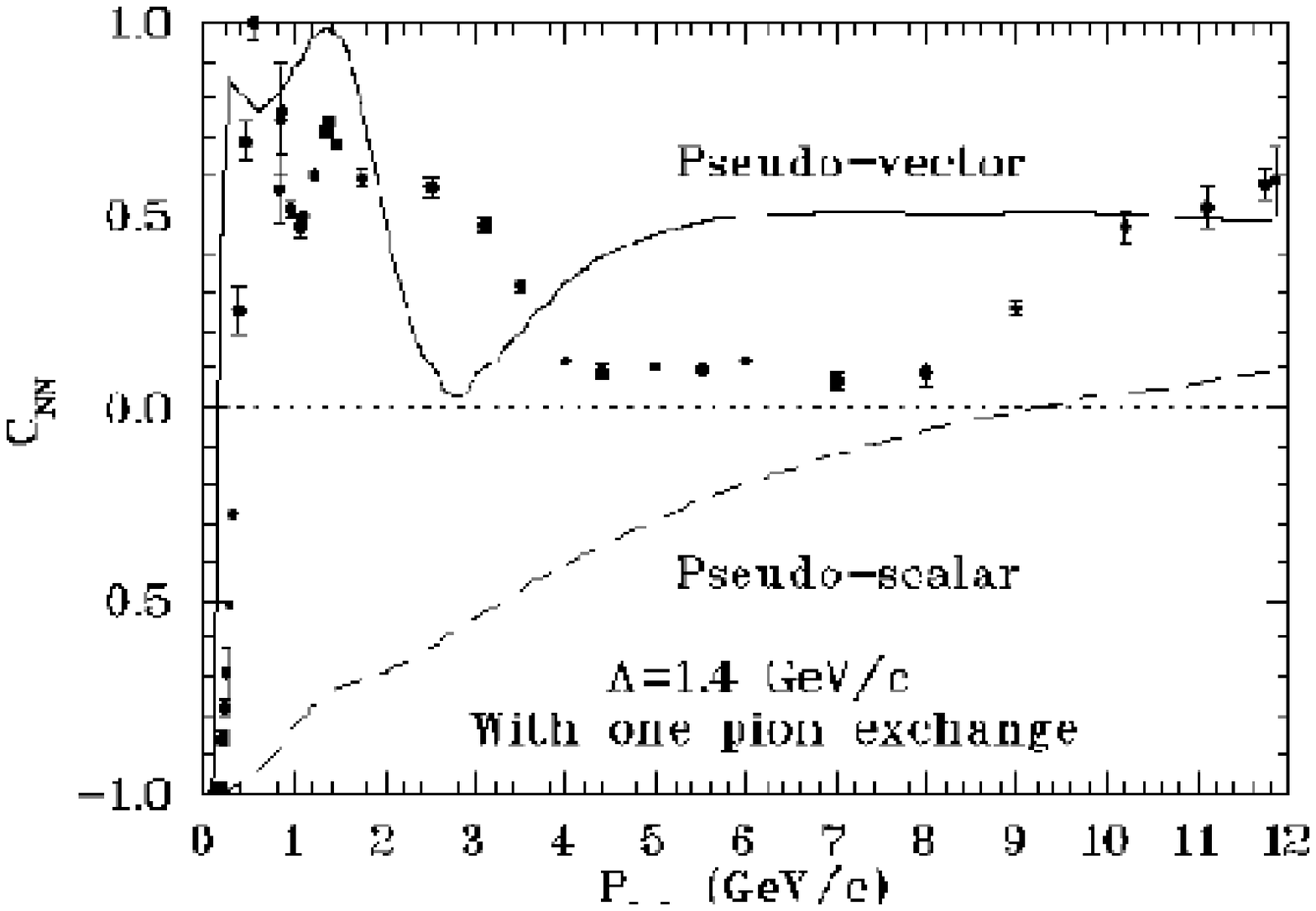,
width=.45\textwidth,silent=,clip=}}
\caption{\label{fig4} Left: Dependence of $|e|^2+|d|^2$ on the type of
coupling.
Right: Comparison with the experimental values of $C_{NN}$ for the 
different couplings.
}
\end{figure}

A second generalization of $\bfq^2$ can be obtained by the following
argument.  The form factor must be a function of Lorentz scalars only. 
There are three four-vectors, the initial and final nucleon momenta (k,k') 
and the pion momentum (q) of which only two are independent.  Choose one
of the nucleon momenta (k) and the pion momentum (q). From these we can
construct three scalars $ q^2,\ k^2,\ \ {\rm and}\ \ k\cdot q .$ In order
to be homogeneous in k and q, the only two invariants we can consider are
$(k\cdot q)^2$ and $k^2q^2$.  The linear combination $(k\cdot q)^2-k^2q^2
$ reduces to $m^2\bfq^2$ if the nucleon is on shell at rest.  
We use the generalization

$$
\bfq^2\rightarrow 
Q^2(k,q)\equiv \frac{(k\cdot q)^2-k^2q^2}{m^2},
$$
which has the property $Q^2(k\pm q,q)=Q^2(k,q)$. Either the initial or
final nucleon momentum may be used. A number of authors have used a 
similar form in one way or another\cite{others}. 

One can also be led to this expression by the requirement that
$Q^2(k,q)=Q^2(k',q)$.  Since $k'=k+q$ (for example) this condition is
suggestive of the vector cross product. The four-dimensional cross product
is defined by the use of a totally antisymmetric 4-component tensor,
$\epsilon_{ijk\ell}$. Since we still have only two vectors (say a and b)
the result is a tensor

$$
T_{ij}=\sum_{k,\ell=0,1,2,3}\epsilon_{ijk\ell}a_kb_{\ell},
$$ 
with 6 independent components.  We can separate the components into two
classes: one in which the zero index is free and one in which it is
summed over
\eq
T_{0j}=\sum_{k,\ell=1,2,3}\epsilon_{0jk\ell}a_kb_{\ell}=[\bfa\times\bfb]_j,
\qe \eq T_{ij}=a_0b_k-b_0a_k=[a_0\bfb -b_0\bfa]_k,\ \ \ i,j,k\ \
{\rm cyclic}\neq 0.
\qe 

Contracting this tensor with the metric tensor we find 
\eq 
\h\sum g_{ii'}g_{jj'}T_{ij}T_{i'j'}= \h \sum
g_{ii}g_{jj}T_{ij}T_{ij}=(a_0\bfb-b_0\bfa)^2-(\bfa\times\bfb)^2 
\equiv (a\cdot b)^2-a^2b^2.
\qe 

In general, two invariants are available, $(k\cdot q)^2/m^2,$ which
evaluates to $q_0^2$ in the rest frame of the nucleon and $[(k\cdot
q)^2-k^2q^2]/m^2$, which evaluates to $|\bfq|^2$. We could choose any
combination of $|\bfq|^2$ and $q_0^2$ for the variable in the rest frame. 
Only $[(k\cdot q)^2-k^2q^2]/m^2$ and $q^2$ are independent of which
nucleon momentum ($k$ or $k'$) is used.

By choosing a function independent of $q_0$ in the nucleon rest frame, the
interaction is instantaneous, perhaps a physically reasonable choice since
the valence quarks are always present, hence they do not have a formation
time. For the present calculation we can always choose the nucleon to be one
of the external lines, and hence on shell. We use

\eq 
f(k,q)= f(\bfk ,q_0,\bfq)  =\left[\frac{\Lambda^2}{
(k\cdot q/m)^2-q^2+\Lambda^2}\right]^2.  
\qe
A product of four of these factors will appear, one for each vertex.
For the box diagram we have
$$
f(\bfk,q_0,\bfq)f(-\bfk,q_0,\bfq)f(\bfk',q_0,\bfq')f(-\bfk',q_0,\bfq'),
$$ 
while for the crossed diagram the factors are
$$f(\bfk,q_0,\bfq')f(-\bfk,q_0,\bfq)
f(\bfk',q_0,\bfq)f(-\bfk',q_0,\bfq').$$
The difference in the variables appearing in these expressions has
important consequences for the behavior of the box and crossed
graphs.

The off-shell range in the calculation, $\Lambda$, which corresponds to
the extension of the proton distribution of partons can be evaluated from
other sources.  Coon and Scadron\cite{coon} found $\Lambda$ between 0.8
and 1.0 GeV/c for a monopole form.  To convert to an equivalent dipole
form at low momentum transfer one can multiply by $\sqrt{2}$ giving a
range from 1.13 to 1.41 GeV/c. In lattice QCD calculations Liu, Dong and
Draper\cite{liu} find $\Lambda$ =0.747 GeV/c for a monopole form and
$\Lambda$=1.32 GeV/c for a dipole. 

We use a dipole form and most of the calculations shown will be for
$\Lambda$=1.4 GeV/c. 

\section{Results}

To describe nucleon-nucleon scattering we use the amplitudes defined by
the Saclay group\cite{saclay} given by the equation
$$
M(\bfk_f,\bfk_i)=\frac{1}{2}[(a+b)+(a-b)
\mbox{\boldmath $\sigma_1$}\cdot\bfn \mbox{\boldmath $\sigma_2$}\cdot\bfn
+(c+d)\mbox{\boldmath $\sigma_1$}\cdot\bfm \mbox{\boldmath $\sigma_2$}\cdot\bfm 
+(c-d)\mbox{\boldmath $\sigma_1$}\cdot\bfl \mbox{\boldmath $\sigma_2$}\cdot\bfl
+e(\mbox{\boldmath $\sigma_1$}+\mbox{\boldmath $\sigma_2$})\cdot\bfn]
$$
\cen{with $\bfl=\frac{\bfk_f+\bfk_i}{|\bfk_f+\bfk_i|},\ 
\bfm=\frac{\bfk_f-\bfk_i}{|\bfk_f-\bfk_i|},\ 
\bfn=\frac{\bfk_i\times\bfk_f}{|\bfk_i\times\bfk_f|}\ .$} 
With identical particle symmetry
$a(90^{\circ})=0$ and  $c(90^{\circ})=-b(90^{\circ})$, so only 3
amplitudes are needed to describe the scattering.
At $90^{\circ}$
\eq 
\sigma=\frac{1}{2}(2|b|^2+|d|^2+|e|^2);\ \  
\sigma C_{NN}=\frac{1}{2}(-2|b|^2+|d|^2+|e|^2),
\qe
so from the data we can extract $2|b|^2$ and $|d|^2+|e|^2$ directly.

Figure 1 shows the amplitudes coming from different types of contribution
for pseudo-scalar coupling. The case of pseudo-vector coupling is
similar. We see that the important contributors are different at
high and low energies. Figure 2 shows the results of the box diagram for
PS and PV coupling for
the partial cross section $|b|^2$.  While the contribution of this diagram
falls rapidly, the crossed diagram falls much more slowly as one can see
from Figure 3 which shows the sum for various values of $\Lambda$. Clearly
the crossed diagram dominates at high energy. This difference can be
traced to the difference in the variables in the form factors as mentioned
above.  We also see a very large sensitivity to the value of the off-shell
range which is very natural since the fall-off of the cross section is
given primarily by the form factor. 

Figure 4 (left) shows the result of the sum of the box and crossed
diagrams for the sum $d^2+e^2$. 

We see that each case there is a large difference between PS and PV
coupling, as pointed out (at lower energies) by Robilotta et al.\cite{rob}. 
However, the PS and PV couplings can be mixed. Gross et al.\cite{mix1}
found 1/4 PS and 3/4 PV. Goudsmit et al.\cite{mix2} found a small mixture
of PS (about 3\%).  

Kondratyuk and Scholten\cite{kon} found a mixture which varied with
momentum transfer, being dominated by PV at low values and about equal at
higher values. This is expected since chiral symmetry imposes
pseudo-vector coupling a low energy. 

Figure 4 (right) shows a comparison with the spin correlation observable.
We see that the PS and PV results bracket the data.

The underlying structure of the proton, represented here by the
form factor, plays an essential role in the scattering in this
energy range.

Our conclusions are: 1) two pion exchange gives a significant contribution
to pp scattering in this energy region,
2) the crossed graph dominates over the box at high energies 
3) the important contributions at high energies are different than
at low energies and 4) there is a significant reduction of the cross
section for PV coupling compared to PS coupling similar to that seen in
low-energy pion-nucleon scattering.
\acknowledgments{The LPNHE is a Unit\'e de Recherche des Universit\'es
Paris 6 et Paris 7, associ\'ee au CNRS.  This work was supported by the U.
S. Department of Energy and the National Science Foundation.}

\end{document}